\documentclass[twoside, prb, english]{revtex4}

\usepackage{makeidx}   % allows index generation
\usepackage{graphicx}  % standard LaTeX graphics tool
\begin{document}

\title{Decoherence of fermions subject to a quantum bath}

\author{Florian Marquardt}

\affiliation{Sektion Physik, Center for NanoScience, and Arnold-Sommerfeld-Center for Theoretical Physics, Ludwig-Maximilians-Universit\"at M\"unchen, Theresienstr. 37, 80333 M\"unchen, Germany}

\email{Florian.Marquardt@physik.lmu.de}

\begin{abstract}
The destruction of quantum-mechanical phase coherence by a fluctuating
quantum bath has been investigated mostly for a single particle. However,
for electronic transport through disordered samples and mesoscopic
interference setups, we have to treat a many-fermion system subject
to a quantum bath. Here, we review a novel technique for treating
this situation in the case of ballistic interferometers, and discuss
its application to the electronic Mach-Zehnder setup. We use the results
to bring out the main features of decoherence in a many-fermion system,
and briefly discuss the same ideas in the context of weak localization. 
[to be published in: ``Advances in Solid State Physics'' Vol. 46, ed. R. Haug, Springer 2006]
\end{abstract}

\maketitle

\section{Introduction}

There are two main messages of this brief review. First, regarding
the physics of decoherence, we will argue that decoherence processes
depend strongly on the type of system (single particle vs. many particles)
and the type of noise (classical vs. quantum). Electronic interference
experiments at low temperatures require a treatment of a many-fermion
system coupled to a quantum bath. In that case, true many-body features
come into play. This includes, in particular, the influence of Pauli
blocking (that tends to restrain decoherence), and the fact that both
hole- and particle-scattering processes contribute equally to the
full decoherence rate. Second, regarding theoretical methods, we review
a novel technique for treating ballistic interferometers subject to
a quantum bath, which is based on the ideas behind the quantum Langevin
equation (as it is known for the Caldeira-Leggett model). This is
more efficient than generic methods (like Keldysh diagrams), and we
will discuss the physical meaning of its ingredients. We apply it
to the interference contrast and the current noise in an electronic
Mach-Zehnder interferometer. We will also mention how the same ideas
(if not the same technical methods) help to understand decoherence
in weak localization within a path-integral framework.

The reasons for studying decoherence range from fundamental aspects
of quantum mechanics to possible applications. On the fundamental
level, the transition from the quantum world (with interference effects)
to the classical world is due to the unavoidable fluctuations of the
environment that tend to destroy macroscopic superpositions very rapidly
(see e.g. \cite{Joos,zurek}). These issues have been studied mostly
with simplified single-particle models of decoherence.

In solid-state physics, decoherence first was investigated in the
field of spin resonance, where straightforward Markoff master equation
treatments are often sufficient \cite{blum,Weiss}. The quantum dissipative
two-level system (spin-boson model) was studied in more detail at
the beginning of the eighties\cite{reviewTwoState}. During the preceding
decade, interest in these questions has seen a revival due to the
prospect of quantum information applications\cite{bennettDiVincenzo},
where the decoherence time has to be at least ten thousand times longer
than the time of elementary operations in order for error correction
to work.

Regarding electronic transport phenomena, which will be our focus
in the following, decoherence effects became important for the first
time during the study of interference effects in disordered conductors,
such as universal conductance fluctuations and weak localization (for
a review see e.g. \cite{beenakkerVanHoutenReview,ImryBook}). Later
on, man-made interference structures were produced in metals and semiconductors,
including Aharonov-Bohm rings, double quantum dot interferometers,
and (most recently) Mach-Zehnder interferometers. These setups are
also important in the quantum information context, both for generating,
transporting, or detecting entanglement, and as highly sensitive measurement
devices. The main nontrivial dependence of the interference contrast
on temperature or transport voltage is produced by decoherence.

\section{Single particle decoherence}

\begin{figure}
\includegraphics[%
  width=1\columnwidth]{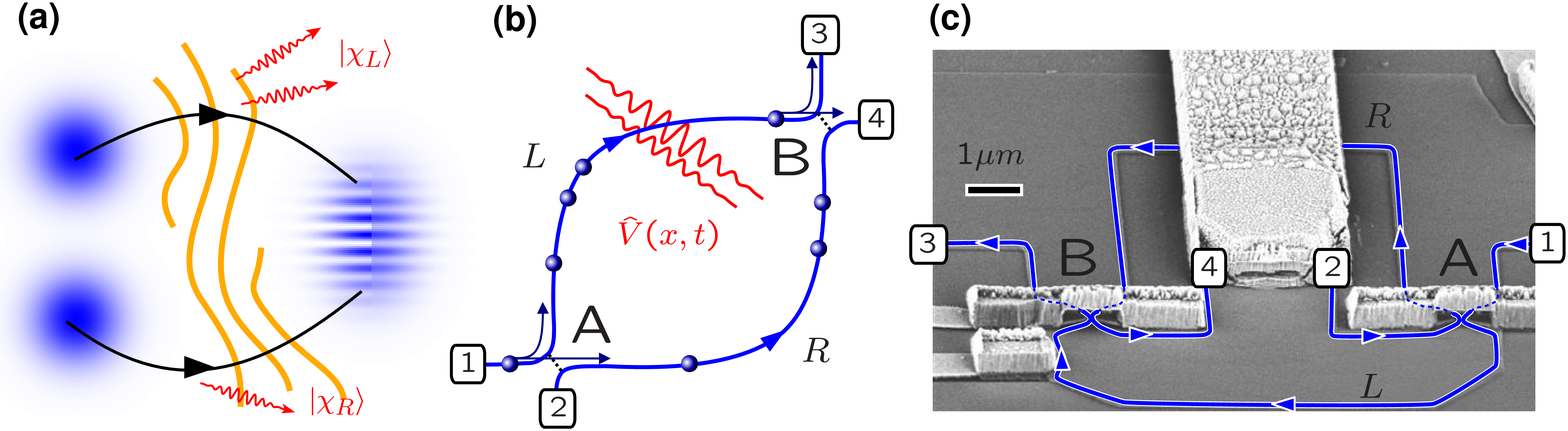}

\caption{\label{cap:(a)-Decoherence-in}(\textbf{a}) Decoherence in a two-way
interferometer. The fluctuations (wiggly lines) introduce a random
relative phase between the two paths. The quantum bath is left behind
in two different states. This blurs the interference pattern (right
part of pattern). (\textbf{b}) Schematic Mach-Zehnder setup. (\textbf{c})
SEM picture of electronic QHE Mach-Zehnder at the Weizmann institute
(courtesy of I. Neder and M. Heiblum)}
\end{figure}

Let us look at a a single particle traversing a two-way interferometer
(Fig. \ref{cap:(a)-Decoherence-in}). Its wavepacket has been split
into two packets $\psi_{L/R}$, going along the two arms (left/right).
After these packets recombine, they form an interference pattern.
This consists of a classical part (sum of probabilities) and an interference
term, which is sensitive to a relative phase:

\begin{equation}
|\psi(x)|^{2}=|\psi_{L}(x)|^{2}+|\psi_{R}(x)|^{2}+\psi_{L}^{*}(x)\psi_{R}(x)e^{i\varphi}+c.c.\end{equation}
What happens once the particle is subjected to (classical) noise,
i.e. a fluctuating potential $V(x,t)$? Even before acceleration/deceleration
effects are noticeable, a random relative phase $\varphi$ between
the two paths is introduced. The actual pattern is obtained by averaging
$|\psi|^{2}$ over many experimental runs. Since $\left|\left\langle e^{i\varphi}\right\rangle \right|\leq1$,
the interference term is suppressed.

At low temperatures ($k_{B}T<\hbar\omega$) we have to consider a
quantum bath, for which there exists an alternative description of
decoherence: The bath acts as a kind of which-way detector, with its
initial state evolving towards either one of two states, $\left|\chi_{R}\right\rangle $
or $\left|\chi_{L}\right\rangle $, depending on the path of the particle
\cite{SAI,lossmullen,ImryBook}. Now it is the overlap of these bath
states that determines the suppression of the interference term. That
overlap is nothing but the Feynman-Vernon influence functional \cite{feynmanvernon}. 

\begin{figure}
\begin{center}\includegraphics[%
  width=0.8\columnwidth]{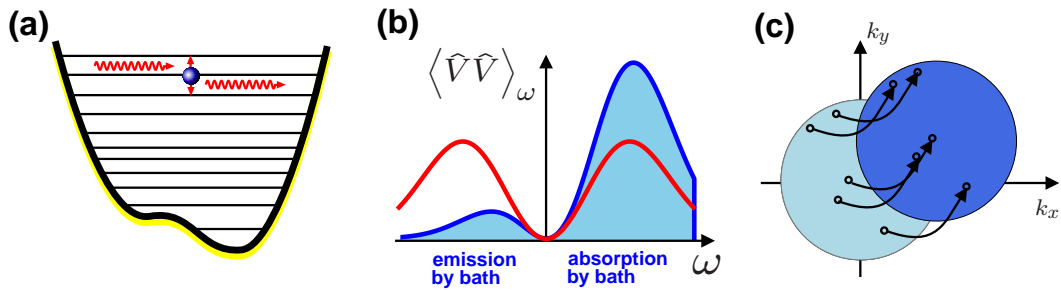}\end{center}

\caption{\label{OverviewCases}(\textbf{a}) The scattering rates $\Gamma_{\downarrow/\uparrow}$
of a highly excited particle are related to the quantum noise spectrum
(\textbf{b}). (\textbf{c}) A many-fermion system subject to classical
noise.}
\end{figure}

Even if the particle is coupled to a quantum bath, decoherence may
still be described using a classical noise spectrum, if the particle's
energy is high and its motion is semiclassical (Fig. \ref{OverviewCases}).
To understand this, consider a simple weak coupling situation, where
the total decoherence rate is given by the sum of downward and upward
scattering rates, calculated using Fermi's Golden Rule. These rates
can be related to the spectrum $\left\langle \hat{V}\hat{V}\right\rangle _{\omega}\equiv\int dte^{i\omega t}\left\langle \hat{V}(t)\hat{V}(0)\right\rangle $
of the quantum noise potential $\hat{V}$. We have $\Gamma_{\downarrow}\propto\left\langle \hat{V}\hat{V}\right\rangle _{\omega}\propto n(\omega)+1$
and $\Gamma_{\uparrow}\propto\left\langle \hat{V}\hat{V}\right\rangle _{-\omega}\propto n(\omega)$,
where $\omega$ is the frequency transfer and $n(\omega)$ the thermal
occupation. Obviously, the sum of these rates does not change if we
replace $\hat{V}$ by classical noise with a symmetrized correlator
$\left\langle VV\right\rangle _{\omega}=(\left\langle \hat{V}\hat{V}\right\rangle _{\omega}+\left\langle \hat{V}\hat{V}\right\rangle _{-\omega})/2$
(red curve in Fig. \ref{OverviewCases} b). This can also be seen
in a more general treatment, using a semiclassical evaluation of the
Feynman-Vernon influence functional. We note that such a replacement
is impermissible near the ground state of the system, where downward
transitions are blocked.%

\section{Many particles}

Up to now, we have considered a single particle subject to classical
or quantum noise. This has been the mainstay of research in quantum
dissipative systems for a long time, with paradigmatic models such
as the spin-boson model or the Caldeira-Leggett model of a single
particle coupled to a bath of harmonic oscillators. However, in electronic
transport experiments (and other setups, e.g. cold atom BEC interferometers)
we are invariably dealing with a many-particle system. What are the
new features arising in that case?

Everything remains straightforward if the noise is classical. Then,
the many-particle problem reduces to the single-particle case: The
wave function of each particle evolves according to the single-particle
Schr\"odinger equation with a given noise field $V(x,t)$. In the
case of many fermions, all the single-particle wave functions remain
orthogonal, forming a Slater determinant (in the absence of intrinsic
interactions). Pauli blocking is then completely unimportant.

We now discuss the one remaining combination: a many-fermion system
coupled to a quantum bath. Unlike all the previous cases, this cannot
be reduced to {}``single particle + classical noise'': True many-body
effects come into play (and appropriate methods are needed). Up to
now, comparatively few quantum-dissipative many-particle systems have
been studied. Examples include open Luttinger liquids \cite{OpenLuttLiquids},
many-electron Aharonov-Bohm rings subject to quantum charge \cite{cedraschi}
or flux \cite{ABRing} noise, many-fermion generalizations of the
Caldeira-Leggett model \cite{ABRing,PRL_DFS,PRA_DFS}, and double
quantum dot interferometers coupled to a quantum bath \cite{DoubleDot}.
Here, we are going to review a recently developed general method of
solution for ballistic interferometers \cite{EPL,MarquardtCondMat},
and then briefly discuss the same physics in the context of disordered
systems (weak localization).

\section{The Mach-Zehnder interferometer}

The Mach-Zehnder (MZ) interferometer arguably represents the simplest
kind of two-way interference setup (Fig. \ref{cap:(a)-Decoherence-in}).
Tuning the relative phase (via the magnetic flux $\phi$) yields sinusoidal
interference fringes in the currents at the two output ports. 

Recently, this model has been realized in electronic transport experiments.
The group of Moty Heiblum at the Weizmann institute managed to employ
edge channels of the integer quantum Hall effect in a two-dimensional
electron gas to build an ideal MZ setup with single-channel transport
and without backscattering \cite{HeiblumEtAl,NederHeiblum}. The group
measured the decrease of visibility (interference contrast in $I(\phi)$)
as a function of rising temperature and transport voltage. No complete
explanation for the results has been provided up to now, especially
for the oscillations in the visibility \cite{NederHeiblum}. Here,
we will explore the possibility that at least part of the decrease
in visibility is due to decoherence processes. 

The effects of \emph{classical} noise $V(x,t)$ onto a MZ setup have
been studied intensively: The suppression of interference contrast
to lowest order in the noise correlator was first calculated in the
work of Seelig and B\"uttiker \cite{Seelig}. Building on this result,
we treated the model to all orders in the interaction, calculating
both the interference contrast and the effects on the shot noise in
the output port of the interferometer \cite{unserPRL,PRB}. The shot
noise has been measured and suggested as a tool to diagnose different
sources of the loss in visibility\cite{HeiblumEtAl}. These studies
have recently been extended to the full counting statistics \cite{FCSMZ}
and a renewed analysis of the dephasing terminal model \cite{DephasingTerminal}.
However, as pointed out above, the situation is more involved for
\emph{quantum} noise, which is needed to account for the loss of visibility
with rising bias voltage.

\section{Equations of motion approach to decoherence in ballistic interferometers}

Recently, we have introduced a novel equations of motion technique
for a many-particle system subject to a quantum bath \cite{EPL},
inside a ballistic interferometer (a detailed discussion may be found
in \cite{MarquardtCondMat}). It is similar in spirit to the quantum
Langevin equation that can be employed to solve the Caldeira-Leggett
model \cite{callegg,Weiss}. Briefly, the idea of the latter is the
following (when formulated on the level of Heisenberg equations).
The total quantum force $\hat{F}$ acting on the given particle, due
to the bath particles, can be decomposed into two parts: 

\begin{equation}
\hat{F}(t)=\hat{F}_{(0)}(t)+\int_{-\infty}^{t}D^{R}(t-t')\hat{x}(t')dt'\label{ForceEquation}\end{equation}
The first describes the intrinsic fluctuations. It derives from the
solution to the free equations of motion of the bath oscillators,
with thermal and quantum (zero-point) fluctuations due to the stochastic
initial conditions. The second part of the force is due to the response
of the bath to the particle's motion. We will call it the {}``back-action''
term, and it gives rise to features such as mass renormalization and
friction. The equation (\ref{ForceEquation}) is valid on the operator
level (not only for averages). In this way, one has {}``integrated
out'' the bath by solving for its motion. Plugging the force $\hat{F}$
into the right-hand-side (rhs) of the Heisenberg equation of motion
for $\hat{x}$ yields the quantum Langevin equation, which in practice
can only be solved for a free particle or a harmonic oscillator (linear
equations). 

\begin{figure}
\includegraphics[%
  width=1\columnwidth]{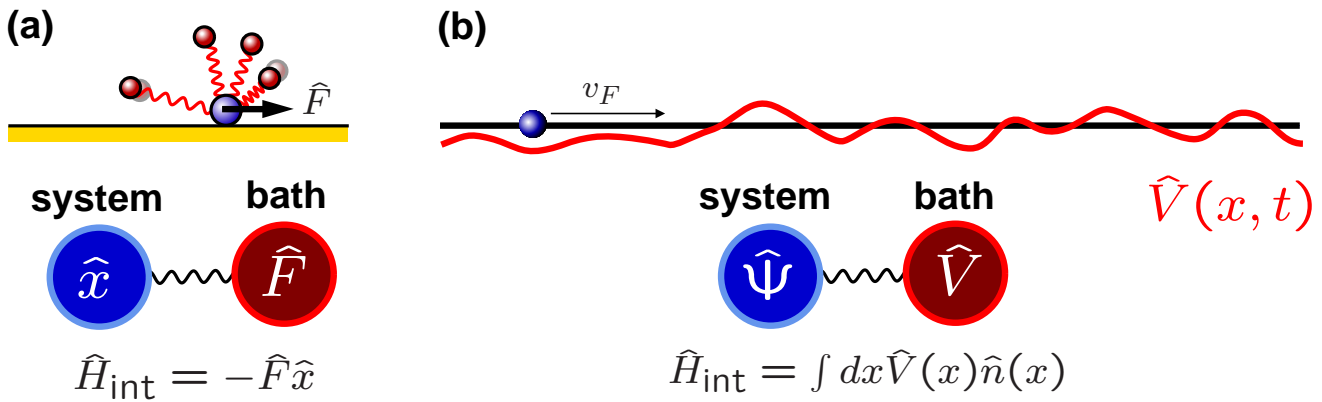}

\caption{\label{cap:Quantum-Langevin-approach}(\textbf{a}) The Caldeira-Leggett
model (single particle and oscillator bath) and (\textbf{b}) a ballistic
many-particle system subject to a quantum noise potential $\hat{V}(x,t)$. }
\end{figure}

In the case of a many-particle system, it is the density $\hat{n}(x)=\hat{\psi}^{\dagger}(x)\hat{\psi}(x)$
that couples to a scalar noise potential $\hat{V}(x)$. The place
of $\hat{x}$ and $\hat{F}$ in the quantum Langevin equation for
a single particle is thus taken by the particle field $\hat{\psi}$
and $\hat{V}$, respectively. Let us now specialize to the case of
fermions traveling ballistically inside the arm of an interferometer.
We will assume chiral motion and use a linearized dispersion relation,
as this is sufficient to describe decoherence. Then the fermion field
obeys the following equation (with a slight approximation \cite{EPL,MarquardtCondMat};
we set $\hbar=1$):

\begin{equation}
i(\partial_{t}+v_{F}\partial_{x})\hat{\psi}(x,t)=\hat{V}(x,t)\hat{\psi}(x,t)\label{PsiEqMotion}\end{equation}
 The formal solution of this equation is straightforward and analogous
to the version for classical noise $V(x,t)$. The particle picks up
a fluctuating {}``quantum phase'' inside a time-ordered exponential:

\begin{eqnarray}
\hat{\psi}(x,t) & = & \hat{T}\exp\left[-i\int_{t_{0}}^{t}dt_{1}\,\hat{V}(x-v_{F}(t-t_{1}),t_{1})\right]\times\nonumber \\
 &  & \hat{\psi}(x-v_{F}(t-t_{0}),t_{0})\,.\label{SolvedPsiEq}\end{eqnarray}
In contrast to the case of classical noise, the field $\hat{V}$ contains
the response to the fermion density, in addition to the intrinsic
fluctuations $\hat{V}_{(0)}$: 

\begin{equation}
\hat{V}(x,t)=\hat{V}_{(0)}(x,t)+\int_{-\infty}^{t}dt'\, D^{R}(x,t,x',t')\hat{n}(x',t')\,.\label{SolvedVEq}\end{equation}
Here $D^{R}$ is the unperturbed retarded bath Green's function, $D^{R}(1,2)\equiv-i\theta(t_{1}-t_{2})\left\langle [\hat{V}(1),\hat{V}(2)]\right\rangle $,
where $\hat{V}$-correlators refer to the free field. With these two
equations, it becomes possible to calculate correlators of the fermion
field (such as current and shot noise).

\section{Decoherence rate in a many-fermion system}

Employing the formal solution from above (and using a lowest-order
Markoff approximation \cite{EPL,MarquardtCondMat}), we find that
the contribution of each electron to the interference term in the
current is multiplied by a factor

$ $\begin{equation}
1-\Gamma_{\varphi}(\epsilon)\tau+i\delta\bar{\varphi}(\epsilon),\end{equation}
with a phase shift $\delta\bar{\varphi}\propto\tau$. We focus on
the suppression brought about by a decoherence rate $\Gamma_{\varphi}(\epsilon)$
that depends on the energy $\epsilon(k)$ of the incoming electron:

\begin{equation}
\Gamma_{\varphi}(\epsilon)=\int_{0}^{\infty}\frac{d\omega}{v_{F}}{\rm DOS}_{q}(\omega)[\begin{array}[t]{c}
\underbrace{2n(\omega)+1}\\
{\rm thermal\,\&\, zeropoint}\\
{\rm fluctuations}\end{array}-\begin{array}[t]{c}
\underbrace{(\bar{f}(\epsilon-\omega)-\bar{f}(\epsilon+\omega))}\\
{\rm from\,"back-action"}\\
\Rightarrow{\rm Pauli\: blocking}\end{array}]\label{GammaPhi}\end{equation}
The rate is an integral over all possible energy transfers $\omega$.
They are weighted by the bath spectral {}``density of states'' ${\rm DOS}_{q}(\omega)=-{\rm Im}D_{q}^{R}(\omega)/\pi$,
where $q=\omega/v_{F}$ for ballistic motion. The first term in brackets
stems from the $\hat{V}_{(0)}$ in the quantum phase. By itself, this
would give rise to an energy-independent rate and a visibility independent
of bias voltage (in contrast to experimental results \cite{HeiblumEtAl,NederHeiblum}).
Thus, the second term is crucially important: It contains the average
nonequilibrium distribution $\bar{f}=(f_{L}+f_{R})/2$ inside the
arms (for equal coupling to both arms) and implements the physics
of Pauli blocking. At $T=0$, it suppresses all transitions that would
take the electron into an occupied state (when $\bar{f}(\epsilon-\omega)=1$
and this cancels against the $1$ from the zero-point fluctuations). 

\begin{figure}
\includegraphics[%
  width=1\columnwidth]{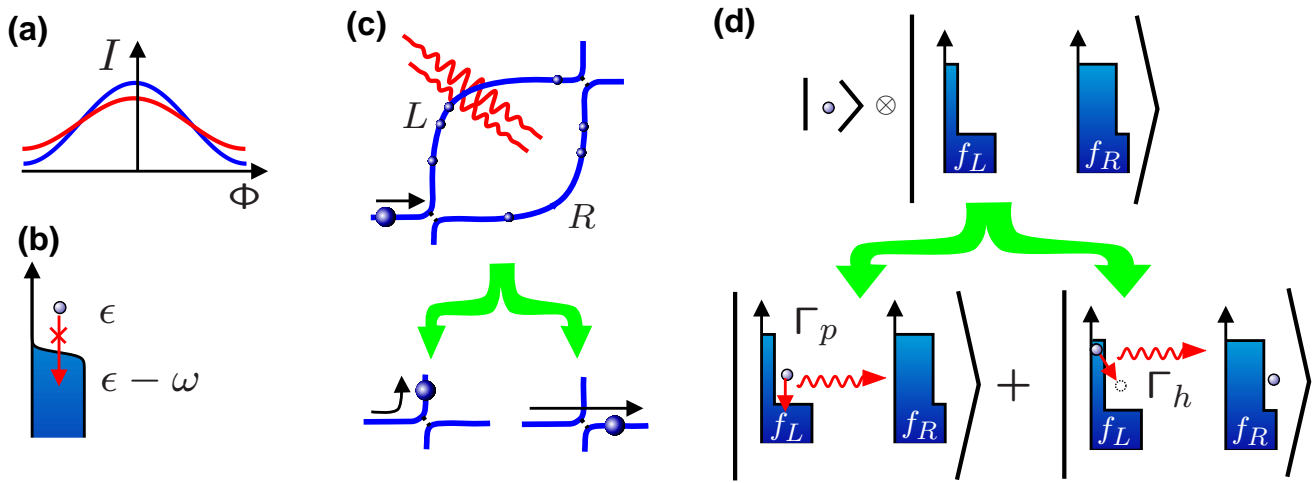}

\caption{\label{DecoherenceInterp}(\textbf{a}) Decoherence suppresses the
visibility of the interference pattern. (red curve) (\textbf{b}) Pauli
blocking restrains decoherence. (\textbf{c}) A particle arriving at
the first beam splitter turns into a coherent superposition $t\left|R\right\rangle +r\left|L\right\rangle $.
(\textbf{d}) For a many-fermion system, both particle- and hole-scattering
contribute to decoherence: $ $ $\Gamma_{\varphi}=(\Gamma_{p}+\Gamma_{h})/2$
.}
\end{figure}

The other main difference (vs. the case of a single particle) is less
obvious but equally important. The decoherence rate $\Gamma_{\varphi}$
is not simply given by the particle-scattering rate, but contains
a contribution from hole scattering processes, where a particle at
another energy $\epsilon+\omega$ is scattered into the given state
at $\epsilon$ (with a factor $f(\epsilon+\omega)$ associated). This
is a generic feature for decoherence of fermionic systems coupled
to a quantum bath, and we now discuss the physical reason. In a single-particle
language, the first beam splitter creates a superposition of the form
$t\left|R\right\rangle +r\left|L\right\rangle $, with $t/r$ transmission
and reflection amplitudes and $R/L$ a packet inside the right/left
arm. In the presence of a sea of other fermions, we should write instead
a superposition of many-body states, for example:

\begin{equation}
t\left|1,1,0,\underline{0},0,0;\,1,1,1,\underline{1},1,0\right\rangle +r\left|1,1,0,\underline{1},0,0;\,1,1,1,\underline{0},1,0\right\rangle \end{equation}
Here the occupations $\left|{\rm left};\,{\rm right}\right\rangle $
of single-particle states in both arms are indicated, with a bar denoting
the energy level $\epsilon$ of interest and the remaining particles
(in the nonequilibrium distributions) playing the role of spectators.
The interference is sensitive to the coherence $t\left|\ldots,\underline{0},\ldots;\,\ldots,\underline{1},\ldots\right\rangle +r\left|\ldots,\underline{1},\ldots;\,\ldots,\underline{0},\ldots\right\rangle $
that requires not only the presence of a particle in one arm but also
the absence of a particle in the other respective arm. This is why
the many-body superposition can equally be destroyed by particle-
and hole-scattering (leading to states with $\left|\ldots,\underline{0},\ldots;\,\ldots,\underline{0},\ldots\right\rangle $
or $ $$\left|\ldots,\underline{1},\ldots;\,\ldots,\underline{1},\ldots\right\rangle $,
respectively).

\begin{figure}
\includegraphics[%
  width=1\columnwidth]{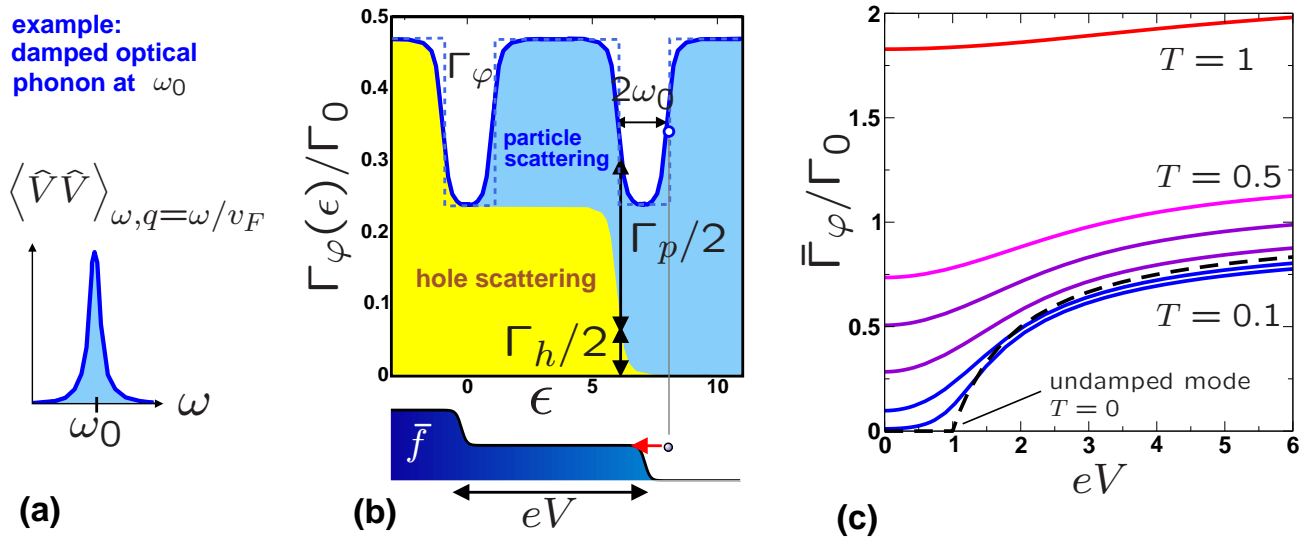}

\caption{\label{cap:The-decoherence-rate}The decoherence rate $\Gamma_{\varphi}$
for the illustrative example of an optical phonon mode (\textbf{a}),
as a function of energy of the incoming electron (\textbf{b}), and
(\textbf{c}) the energy-averaged rate $\bar{\Gamma}_{\varphi}$ as
a function of voltage $V$ and temperature $T$ (in units of $\omega_{0}$).
Dashed curves in (\textbf{b}) and (\textbf{c}) refer to an ideal undamped
mode at $T=0$.}
\end{figure}

We have illustrated this in Fig. \ref{cap:The-decoherence-rate},
where we have chosen a simple model bath spectrum (a broadened optical
phonon mode at $\omega_{0}$). Let us focus on small temperatures
$T\ll\omega_{0}$. If the electron is far above the Fermi sea, it
can easily undergo spontaneous emission and lose its coherence, thus
the decoherence rate is maximal. For smaller energies, near the upper
step in $\bar{f}$, it might end up in an occupied state, thus $\Gamma_{p}$
is reduced, leading to a dip in $\Gamma_{\varphi}$. Inside the transport
voltage window, $\Gamma_{p}$ remains at $1/2$ its previous value,
but now $\Gamma_{h}$ raises $\Gamma_{\varphi}$ back to its maximal
value. Finally, another dip is observed near the lower edge of the
voltage window. The visibility is directly given by $1-\bar{\Gamma}_{\varphi}$,
with $\bar{\Gamma}_{\varphi}$ the energy-average of $\Gamma_{\varphi}$
over the voltage window. At $T,V\rightarrow0$, $\bar{\Gamma}_{\varphi}$
vanishes. Decoherence sets in only when the electron can emit phonons
and thereby reveal its path through the MZ setup. At higher temperatures,
the Fermi distribution becomes smeared, thereby easing the restrictions
of Pauli blocking, and the thermal fluctuations of the bath grow,
increasing $\Gamma_{\varphi}$. For $T\gg\omega_{0}$, the energy/voltage-dependence
of $\Gamma_{\varphi}$ becomes unimportant, and an approximate treatment
becomes possible, replacing the quantum bath by classical noise.

\section{Decoherence in weak localization}

We now briefly discuss how the same concepts apply to weak localization
\cite{WLdephasingClassical,WLdephasingDiags,ChakSchmid,beenakkerVanHoutenReview},
where the constructive interference of time-reversed pairs of diffusive
trajectories increases the electrical resistance of a disordered sample.
One is interested in the linear response conductance, where the external
perturbation (the electric field) induces a particle-hole excitation
by lifting one of the particles above the Fermi sea. This creates
a many-body state similar to the one above, $\sqrt{1-\delta^{2}}\left|\ldots,\underline{1},\ldots,\underline{0},\ldots\right\rangle +\delta\left|\ldots,\underline{0},\ldots,\underline{1},\ldots\right\rangle $,
where $\delta$ is the small amplitude of the excited state. Following
arguments analogous to those above \cite{JvDAndMe}, we see again
why both particle- and hole-scattering processes contribute to the
decoherence rate. 

Many discussions of decoherence in weak localization have focussed
on the thermal (classical) part of the Nyquist noise \cite{WLdephasingClassical,ChakSchmid}.
This leads to a single-particle problem that can be treated using
path integrals \cite{ChakSchmid}. Diagrammatic calculations of the
decoherence rate in the presence of a quantum bath \cite{WLdephasingDiags}
yielded results that can be interpreted in the manner discussed above.
It is obviously desirable, though difficult, to cast these as well
into the powerful path-integral framework. Golubev and Zaikin were
the first to present a formally exact influence functional approach
for many-fermion systems \cite{FV-Fermi}. Their semiclassical evaluation
yielded a decoherence rate that does not vanish at $T=0$ and is independent
of electron energy, in contrast to diagrammatic calculations and the
ideas about Pauli blocking discussed above. 

Recently, we have revisited this problem \cite{JvDAndMe,BetheSalpeterAndJan}
and have shown that the results of much more complicated diagrammatic
calculations \cite{aleineraltshuler,BetheSalpeterAndJan} can be exactly
reproduced by a rather simple prescription. The case {}``many particles
+ quantum bath'' may be reduced to {}``single particle + classical
noise'', provided one uses an effective, modified noise spectrum
of the following form\cite{JvDAndMe}:

\begin{equation}
\left\langle \hat{V}\hat{V}\right\rangle _{\omega}\mapsto\left\langle VV\right\rangle _{\omega}^{{\rm eff}}\equiv\frac{1}{2}\left\langle \left\{ \hat{V},\hat{V}\right\} \right\rangle_{\omega} +\frac{1}{2}\left\langle \left[\hat{V},\hat{V}\right]\right\rangle_{\omega} (f(\epsilon+\omega)-f(\epsilon-\omega))\end{equation}
The first part is the symmetrized quantum correlator, containing the
zero-point fluctuations. The second part incorporates Pauli blocking.
These terms correspond to those in the equation of motion approach,
Eq. (\ref{GammaPhi}), with which this method is consistent. The resulting
decoherence rate vanishes at $T=0$ \cite{GZcomment}. Earlier similar
ideas \cite{Eiler,ChakSchmid,CohenImry} represent approximations
to the present approach.

\begin{figure}
\includegraphics[%
  width=1\columnwidth]{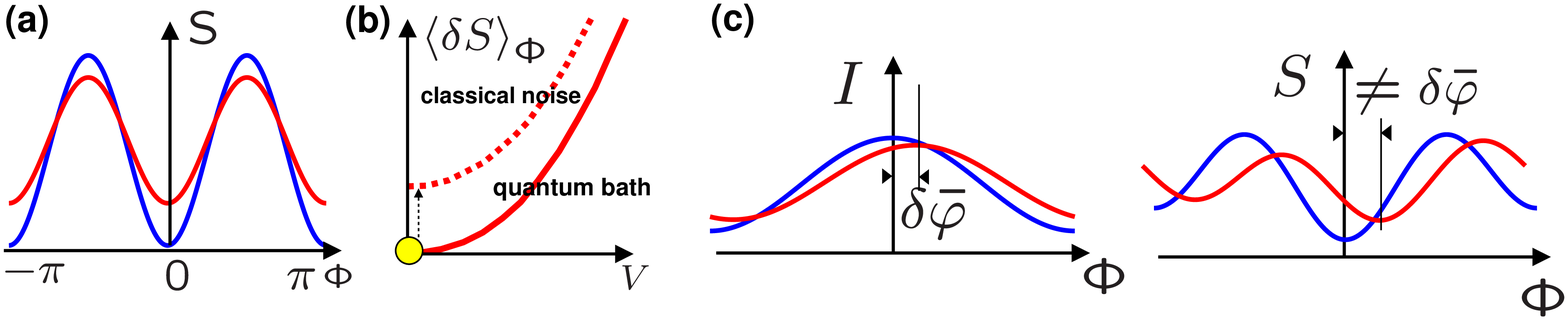}

\caption{\label{ShotNoise}Main effects of a quantum bath on shot noise in
a MZ setup: (\textbf{a}) Suppression of visibility in $S(\phi)$.
(\textbf{b}) No Nyquist noise correction, but classical conductance
fluctuations $S\propto V^{2}$ at high voltages. (\textbf{c}) Different
phase shifts in $I(\phi)$ and $S(\phi)$ for asymmetric setups.}
\end{figure}

\section{Effects of a quantum bath on shot noise}

We briefly return to the MZ setup. Using our approach, it is possible
to discuss the influence of the quantum bath on the shot noise \cite{BlanterBuettiker}
power $S$ in the output port \cite{EPL,MarquardtCondMat}. The visibility
of the interference pattern $S(\phi)$ is reduced, although this cannot
be described by the same decoherence rate as for the current. The
most important feature refers to the phase shifts observed in asymmetric
setups, for the current and the shot noise. These can become different:
$I(\phi)=\bar{I}+\delta I\cos(\phi-\delta\bar{\varphi})$ and $S(\phi)=\bar{S}+S_{1}\cos(\phi-\delta\phi_{1})+S_{2}\cos(2(\phi-\delta\phi_{2}))$,
with $\delta\bar{\varphi}\neq\delta\phi_{1}\neq\delta\phi_{2}$ in
general. This prediction is in contrast to all simpler models (involving
classical noise etc.), which usually do not give rise to phase shifts
at all. Something like this seems to have been observed in recent
experiments at the Weizmann institute. Another equally important avenue
of current research is the application to nonlinear (non-Gaussian)
environments.

\section{Conclusions}

Decoherence in transport interference situations can often be reduced
to the case of a single particle subject to classical noise. However,
for a many-fermion system subject to a quantum bath, true many body
features remain and have to be taken into account via suitable techniques.
The two main physical features are Pauli blocking and the importance of both hole- and particle-scattering
processes. The technical innovation reviewed here is an equation of
motion approach that is well suited to describe decoherence of many
particles moving in ballistic interferometers. We have discussed its
application to the MZ interferometer setup, the loss of visibility
in the current and (briefly) the effects on shot noise. In addition,
we have pointed out that the same kind of physics applies to decoherence
in weak localization.

I thank M. Heiblum, I. Neder, J. v. Delft, C. Bruder, D. S. Golubev,
M. B\"uttiker, Y. Imry, R. Smith, V. Ambegaokar, S. M. Girvin, and all the other
people with whom I discussed and worked on these topics over the years.

\end{document}